\documentstyle[epsfig,aps,prl,twocolumn,doublespace,12pt]{revtex}
\begin{document}
%
%
%
%
\onecolumn
\title{The Pauli Exclusion Principle and $SU(2)$ Versus $SO(3)$ in Loop Quantum
Gravity}
\author{John Swain}
\address{Department of Physics, Northeastern University, Boston, MA 02115, USA\\
email: john.swain@cern.ch}
\date{Submitted for the Gravity Research Foundation Essay Competition, March 27,
2003 where it received an Honourable Mention, and extended for publication July 16, 2003}
\maketitle

\begin{abstract}
\section*{\bf Abstract}

Recent attempts to resolve the ambiguity in the loop quantum
gravity description of the quantization of area has led to 
the idea that $j=1$ edges of spin-networks dominate in 
their contribution to black hole areas as opposed to $j=1/2$ which
would naively be expected. This suggests that
the true gauge group involved might be $SO(3)$ rather than $SU(2)$
with attendant difficulties.
We argue that the assumption that a
version of the Pauli principle is present in loop quantum gravity
allows one to maintain $SU(2)$ as the gauge group while still 
naturally achieving the desired suppression of spin-1/2 punctures.
Areas come from $j=1$ punctures rather than $j=1/2$ punctures
for much the same reason that photons lead to macroscopic classically
observable fields while electrons do not.
\end{abstract}
%
%
\section{Introduction}
The recent successes of the approach to canonical quantum gravity
using the Ashtekar variables have been numerous and significant. Among
them are the proofs that area and volume operators have discrete
spectra, and a derivation of black hole entropy up to an overall
undetermined constant\cite{Ashtekarstuff}. An excellent recent review
leading directly to this paper is by Baez\cite{Baez}, and
its influence on this introduction will be clear.

The basic idea is that a basis for the solution of
the quantum constraint equations is given by {\em spin-network} states,
which are graphs whose edges carry representations $j$ of $SU(2)$.
To a good approximation, the area $A$ of a surface which intersects
a spin network at $i$ edges, each carrying an $SU(2)$ label $j$ is given
in geometrized units (Planck length equal to unity) by
\begin{equation}
A \approx \sum_i 8\pi\gamma\sqrt{j_i(j_i+1)}
\end{equation}
where $\gamma$ is the Immirzi-Barbero parameter\cite{Immirzi-Barbero}.
The most important microstates 
consistent with a given area are those for which $j$ is as small as possible,
which one would expect to be $j_{min}=1/2$. In this case, each contribution
to the area corresponds to a spin $j=1/2$ which can come in two possible
$m$ values of $\pm 1/2$. For $n$ punctures, we have
$A \approx 4\pi\sqrt{3}\gamma n$ and entropy $S\approx \ln(2^n) \approx
\frac{\ln(2)}{4\pi\sqrt{3}\gamma}A$. 

Now looking outside loop quantum gravity for help, we can use
Hawking's formula\cite{Hawking} for black hole entropy $S=A/4$ to get
$\gamma=\frac{\ln(2)}{\pi\sqrt{3}}$ and the smallest quantum of 
area is then $8\pi\gamma\sqrt{\frac{1}{2}(\frac{1}{2}+1)} = 4\ln(2)$.
Physically this is very nice as it says that a black hole's horizon
acquires area, to a good approximation, from the punctures of many
spin network edges, each carrying a quantum of area $4\ln(2)$ and
one ``bit'' of information -- a vindication of Wheeler's ``it from bit''
philosophy\cite{Wheeler}.

Bekenstein's early intuition\cite{Beckenstein} that the area operator
for black holes should have a discrete spectrum made of equal area steps
(something not really quite true in loop quantum gravity in full generality)
was followed by Mukhanov's observation \cite{Mukhanov} that 
the $n^{th}$ area state should have degeneracy $k^n$ with
steps between areas of $4\ln(k)$ for $k$ some integer $\geq 2$ 
in order to reproduce the Hawking expression $S=A/4$. For $k=2$ one would have
the $n^{th}$ area state described by $n$ binary {\em bits}. 

On the other hand,  Hod\cite{Hod} has argued that by looking at the
quasinormal damped modes of a classical back hole one should be
able to derive the quanta of area in a rather different way. The basic
idea is to use the formula $A=16\pi M^2$ relating area and mass of a
black hole to get $\Delta A=32\pi M \Delta M$ for the change in area 
accompanying an emission of energy $\Delta M$. Nollert's computer
calculations\cite{Nollert}
of the asymptotic frequency $\omega$ of the damped normal modes gave
$\omega \approx 0.4371235/M$, so setting $\omega = \Delta M$ one finds
$\Delta A \approx 4.39444$. It is tempting then to conclude that perhaps
$\Delta A = 4\ln(3)$. Motl \cite{Motl} later showed
that this is indeed correct, and not just a fortuitous numerical coincidence.

Since then, Dreyer\cite{Dreyer} has pointed out that one might well
expect $\Delta A \approx 4\ln(3)$ instead of $\Delta A \approx 4\ln(2)$
if the spin network edges contributing to the area of a black
hole didn't carry $j=1/2$, but rather $j=1$. In this case
$j_{min}$ would be $1$ rather than $1/2$,
there would be
three possible $m$ values, and area elements would be described not by
binary ``bits'', but by trinary ``trits''. (See also \cite{MotlandNeitzke}).
This also suggests
that perhaps the correct gauge group is not $SU(2)$ but $SO(3)$, although
this could complicate the inclusion of fermions in the theory.

Corichi has recently argued\cite{Corichi} that one might arrive at
the conclusion that $j_{min}=1$ by suggesting that one should think of
a conserved fermion number being assigned to each spin-1/2 edge.
Adding or losing an edge's worth of area would have to mean
that at some point a spin-1/2 edge would be essentially dangling
in the bulk ({\it i.e.} not imbuing the horizon surface with area)
and this should not be allowed. If edges carried $j=1$ 
one could imagine coupling the edge to a fermion-antifermion pair
and this would locally solve the fermion number problem. This is 
quite appealing as one might then think of the loss of an element of
area with accompanying fermion-antifermion production in Hawking
radiation as the detachment of a spin-1 edge from the horizon
which then couples to an $f\bar{f}$ pair. As Corichi\cite{Corichi} points out:

{\em ``the existence of $j=1/2$ edges puncturing the horizon is not forbidden
\ldots, but they must be suppressed. Thus, one needs a dynamical 
explanation of how exactly the entropy contribution is dominated by
the edges with the dynamical allowed value, namely $j=1$.''}

\section{The Exclusion Principle}

The point of this essay is to suggest that one 
might want to 
assume that
a version of the spin-statistics theorem (or, equivalently, the
Pauli exclusion principle)
applies to loop quantum gravity. More precisely, it could be the
case that no more than two punctures of $j=1/2$, each with differing
$m$ values, may puncture a given surface. In this case,
the dominance of $j=1$ punctures
(even though $j=1/2$ is allowed) is very natural: if only a maximum
of two spin-1/2 edges can puncture any surface then for large
numbers of punctures one would have
an effective $j_{min}=1$ despite the gauge group being $SU(2)$.

It is not immediately obvious what (if any) sort of extension of the exclusion
principle for particles in spacetime carrying $SU(2)$ representation labels
should apply to spin network edges carrying $SU(2)$ representation labels. 
In quantum
mechanics, the spin-statistics theorem is simply a postulate, since,
as Dirac\cite{Dirac} puts it ``to get agreement with experiment one
must {\em assume} that two electrons are never in the same state'' 
(my italics). The
same sort of reasoning could be applied here, but it is also possible
to make a stronger case for the idea. 

The spin-statistics theorem as usually formulated and proven
(to the extent that one rigorously proves anything in quantum field
theory!), is, of course,
for matter fields in a background spacetime usually assumed to be
flat. 
For bosonic fields one makes the usual expansion of plane
waves in terms of creation and annihilation operators and then demands
that at spacelike separations the field operators should commute. This
is certainly physically reasonable and is meant to capture the appropriate
notion of causality, for one would not expect physics at some point in spacetime
to be affected by what goes on outside its light cone. For fermion fields
one is led to look at anticommutators instead in order to reproduce the
experimentally observed fact you don't find two identical fermions in 
the same state. Anticommutators were basically pulled out of
thin air by Jordan\cite{Jordan} and any
relation now to causality is obscured unless one
assumes that the fermion fields are Grassman-valued. Most modern
quantum field
theoretic proofs of the spin-statistics theorem are essentially carried out
by seeing what happens if one assumes 
the wrong choices of commutation or anticommutation relations
and finding that things don't work out. For example, quantizing the Dirac
field with commutators gives an unstable vacuum. A rather comprehensive
review of the history and literature is in \cite{Paulibook}.

Here we don't have particles, but rather edges of spin networks, and
we don't really have spacetime either except in some approximation,
so we have to try to think about the spin-statistics theorem in a slightly
less spacetime-bound way. That being said, the fact that the edges
are meant to puncture a spacelike surface like the horizon of a black hole
is encouraging as it might make some sense to think of the punctures
are being spacelike separated and possibly subject to commutation or
anticommutation relations. In addition, the 
spin-statistics association is strongly
combinatorial in flavour and seems natural in a spin-network context.

For a surface punctured by spin network edges I want to argue that one should
consider an amplitude which returns to its original value, up to 
a phase, upon the exchange of two spin-1/2 (and thus identical,
indistinguishable) punctures. If making the exchange twice leads to 
the identity\footnote{It is interesting to consider the possibility of
more exotic braid or anyon-like statistics if one would have
to keep track of how one edge moved around another, but this is
beyond the scope of this essay, but see \cite{Oeckl}.}, one then needs merely to
choose a sign, and -1 seems at least as natural as +1. This argument
can be sharpened in the following way:

Let us consider the configuation space of $n$ spin-$j$ non-coincident
{\em identical} punctures 
and see what we can expect on fairly general grounds. First of all, 
as shown long ago by Laidlaw and DeWitt\cite{LaidlawDeWitt},
that phases for propagators in a multiply connected 
configuration space must form a scalar unitary representation of
the fundamental group. For us, the configuration space is
the set of $n$ non-coincident points on a sphere, and
this means that the phases must form 
a scalar unitary representation of the permutation group.
That then
limits the possible choice of statistics 
to Bose statistics (no phase change under permutations) 
or Fermi statistics (change of sign for any odd permutation).
As it stands, this is just
a statement about possible statistics and has nothing to do with rotations,
$SU(2)$, $SO(3)$, or even physical space, 
but it does fix the possible choices we can make.

Now to argue for Fermi statistics for odd half-integer $j$ punctures
and Bose statistics for integer $j$ punctures we cannot use the usual
QFT arguments. We have no suitable creation and
annihilation operators, no background spacetime, no plane wave expansion,
{\em etc.} In particular, 
simple arguments based on identifying an exchange as a
composition of physical rotations ({\it i.e.} \cite{Donth}) seem 
inapplicable as we don't have a background space in which to rotate, as do
arguments based on extended kink-like objects ({\it i.e.} \cite{kinks})
since the punctures are meant to be points. Related approaches
such as those of Balachandran {\it et al.} \cite{Balachandran},
Tseuschner\cite{Tscheuschner}, Berry and Robbins\cite{Berry}, and many others
in \cite{Paulibook} all seem difficult to apply here.

What we need to do is look for
a proof (or at least an argument)
for the spin-statistics relation that is not rooted in
a prior concept of physical space. 
One possibly applicable
route is to just go directly to the configuration space and use
ideas from geometric quantization\cite{Geometricq},
which I now do, 
following rather beautiful arguments
of Anastopoulos\cite{Anastopoulos}.

In geometric quantization one starts with a classical phase space
$\Gamma$ with its associated symplectic form $\Omega$.
Quantum mechanics and complex numbers enter
via the {\em prequantization}
of $(\Gamma,\Omega)$ given by a $U(1)$ fiber bundle $(Y,\Gamma,\pi)$ with
total space $Y$, base space $\Gamma$ and projection map $\pi:Y\rightarrow
\Gamma$. A $U(1)$ connection $\omega$ in $Y$ is required to satisfy 
$d\omega = \pi_\ast \Omega$. 
This then requires the integral of
$\Omega$ over any surface
$\int \Omega$ to be a multiple of $2\pi$. (The line
integral of $d\omega$ is clearly zero around any closed path $\gamma$.
This corresponds to trivial holonomy $\exp(i\int_\gamma \omega)$, which
can also be calculated by Stokes theorem from
the integral of $\Omega$ on any surface $\sigma$ bounded by such a closed path
and the result follows.) Sections of the $U(1)$ bundle (suitably completed)
then form the Hilbert space for the quantum system corresponding to 
the classical $(\Gamma,\Omega)$.

Now consider the quantization of the sphere $S^2$, coordinatized
$S^2=\lbrace (x_1,x_2,x_3) | x_1^2+x_2^2+x_3^2=1 \rbrace$
with the symplectic form

\begin{equation}
\Omega = \frac{1}{2} s \epsilon_{ijk} x^i dx^j\wedge dx^k
\end{equation}
and a symplectic action of $SO(3)$ on $S^2$ where $SO(3)$ acts on 
the $x^i$ in the usual way by its defining representation and
obviously leaves $\Omega$ invariant.
Each choice of $s$ gives a different symplectic manifold,
and the requirement that $\Omega$ be
integrable requires that $s=n/2$ with $n$ an integer.
In this way $s$ corresponds to
the usual notion of spin in quantum mechanics.
Note that so far there is no explicit identification of the $x^i$ with
spacetime directions -- they just happen to define the coordinates on
an abstract $S^2$.

An explicit realization
of the $U(1)$ bundle is provided by the Hopf map
$\pi(\xi)^i = \bar{\xi}\sigma^i\xi$
which is defined in terms of 2-component spinors $\xi$
normalized to length 1 by $\bar{\xi}\xi=1$
with $\sigma^i$ the usual
Pauli matrices. Note that the  $\xi$ with this normalization
lie on $S^3$. There is also the natural connection
$\omega=-i\bar{\xi}d\xi$ and a $U(1)$ action on the fibers
which is just multiplication by a phase. This $U(1)$ action will be
very important in what follows.

The Hopf map clearly takes one from the $U(1)$ bundle
over $S^2$ down to $S^2$ since $\pi(\xi)^i$ is obviously real
and invariant under $\xi\rightarrow u\xi$ for any $u=\exp(i\theta)$ in
$U(1)$. While a full treatment for arbitrary $s$ can be
found in \cite{Anastopoulos}, let us just consider the case of
$s=1/2$. In this case we have $U(1)$ operations corresponding
to the two square roots of unity $\pm 1$, both of which correspond
to the identity element of $SO(3)$.

We can now consider the bundle whose total space is the set
of orbits of $S^3$ under the two $U(1)$ actions of multiplication
by $\pm 1$ and the same projection map $\pi$. This is our
prequantization. $SU(2)$ actions 
on $S^3$ correspond to $SO(3)$ actions on $S^2$, and we pick
up a factor of -1 for a $2\pi$ rotation. We can now think
of an $s=1/2$ state as a point on the sphere $S^2$, accompanied
by this sign change for $2\pi$ rotations.
(Much the same argument
goes through in general for higher spins in a similar fashion: for
spin n/2 we have n roots of unity $\exp(2\pi i r/n)$ with $r$ ranging
from 0 to $n-1$ and the orbits of $S^3$ under these $U(1)$ actions
is the prequantization. As one might expect, there is a sign change
under $2\pi$ rotations for half-integer spins and none for integer spins.)

Now consider two classical phase spaces $\Gamma_i$ ($i=1,2$) with symplectic forms
$\Omega_i$. The combined system then has classical phase space
$\Gamma_1 \times \Gamma_2$ and symplectic form $\Omega_1 \oplus \Omega_2$.
Repeating the same construction, and assuming the two $\Gamma_i$ are
the same and both $S^2$, we can ask what it would take to effect an {\em exchange}
carrying $(\xi_1,\xi_2)$ to $(\xi_2,\xi_1)$. This can be accomplished by
two $SO(3)$
rotations, one of the first $\Gamma_1(=S^2)$ on which $\xi_1$ lies, and one
on the second $\Gamma_2(=S^2)$ such that the $\xi_1\rightarrow \xi_2$ and
$\xi_2\rightarrow \xi_1$. If we believe that $\xi_1$ and $\xi_2$ are
{\em indistinguishable}, however, we should really think of them as
two points on the {\em same} $S^2$. That means the rotation that exchanges
them should act along the same orbit of the $SO(3)$ action. The net result
then is a rotation of $2\pi$ and one has a sign of $\pm 1$ depending
on the parity of $n$ just as one would want for a spin-statistics theorem.
In other words, the intuitive arguments referred to earlier which are
used to argue for the spin-statistics theorem following from the equivalence
of an exchange with a rotation of $2\pi$ in physical space can also be
used here even though the $S^2$ was just introduced as a way of dealing with
$SO(3)$ and, through geometric quantization, $SU(2)$.

In other words, I would argue that even though one might colloquially speak
of the edges as carrying ``spins'', knowing full well that this is really
a way of saying ``$SU(2)$ representation labels with no obviously necessary
connection to spin of elementary particles or irreducible representations
of the rotation group in physical space'', in fact it {\em does} make
sense to think of them as physical spins and argue for a
spin-statistics theorem. In this sense the spin-statistics theorem might
better be thought of as a sort of
``($SU(2)$ representation label)-(sign change or not on exchange)''theorem.

In loop quantum gravity this leads
then to a picture in which a reasonably large black hole {\em can} get area
contributions from
spin-1/2 (and spin-3/2, spin-5/2, {\em etc.}) punctures, but these
are always very small compared to the enormous number of $j=1$
edges. The value $j=1$ is the lowest value of $j$ contributing
nonzero area not
being severely limited by Fermi-Dirac statistics, and able to 
appear arbitrarily often. This can then make it look like we're dealing
with $SO(3)$ rather than $SU(2)$.

\section{Conclusions}

In a sense, the question of $SU(2)$ vs. $SO(3)$ in 
loop quantum gravity could be
very much like one that we face in everyday physics. Integer spin particles,
which fall into $SO(3)$ representations, obey Bose-Einstein statistics and
gregariously bunch together to give large macroscopically observable
fields such as electromagnetic fields.  Half-integer spin particles do not.
We could well be excused
for thinking 
that the symmetry group of our world under rotations was $SO(3)$
rather than $SU(2)$. Indeed, until the discovery of spin, it did appear
that physical rotations were always elements of $SO(3)$. The
need for $SU(2)$ was, in many ways, a surprise!

It may be hard to find direct experimental evidence of these ideas,
but it is at least possible to make some predictions.
For example, the $SU(2)$ theory with the exclusion principle proposed
here will give both:

\begin{itemize}
\item[a)] what seems to be the correct result for large
black holes, with areas well-described by values which go up {\em in
steps of} $4\ln(3)$; {\underline{and}}
\item[b)] the possibility of
simultaneously admitting areas {\em as small as} $4\ln(2)$.
\end{itemize}

\section*{Acknowledgements}

I would like to thank the NSF for their continued 
and generous support and an anonymous referee for useful comments
and suggestions.


\begin{thebibliography}{99}
\bibitem{Ashtekarstuff} A. Ashtekar and J. Lewandowski,
{\sl Class. Quant. Grav} {\bf 14} (1997) A55; C. Rovelli and L. Smolin,
Nucl. Phys. {\bf B442} (1995) 593, {\em Erratum-ibid.} {\bf B456} (1995) 753. 
\bibitem{Baez} J. Baez in ``Matters of Gravity, the 
newsletter of the APS Topical Group on Gravitation, Spring 2003'',
J. Pullin (ed.), {arXiv:gr-qc/0303027}.
\bibitem{Immirzi-Barbero} A. Ashtekar {\it et al.},
Phys. Rev. Lett. {\bf 80} (1998) 904;
G.\ Immirzi, 
{\sl Nucl.\ Phys.\ Proc.\ Suppl.\ }{\bf 57} (1997), 65. 
F.\ Barbero, {\sl Phys.\ Rev.\ }{\bf D51} (1995), 5507.
\bibitem{Hawking} S.\ Hawking, {\sl Commun.\ Math.\ Phys.\ } {\bf 43} (1975), 199.
\bibitem{Wheeler} J.\ Wheeler, in {\sl Sakharov 
Memorial Lecture on Physics,} vol.\ 2, eds.\ L.\ Keldysh and V.\ 
Feinberg, Nova Science, New York, 1992.
\bibitem{Beckenstein} J.\ Bekenstein, {\sl Phys.\ Rev.\ }{\bf D7} 
(1973), 2333.
\bibitem{Mukhanov} V.\ Mukhanov, {\sl 
JETP Lett.\ }{\bf 44} (1986), 63; \\
J.\ Bekenstein and V.\ Mukhanov, 
{\sl Phys.\ Lett.\ }{\bf B360} (1995), 7. 
\bibitem{Hod} S.\ Hod,
{\sl Phys.\ Rev.\ Lett.\ }{\bf 81} (1998), 
4293;\\
S.\ Hod, {\sl Gen.\ Rel.\ Grav.\ } {\bf 31} (1999), 1639.
\bibitem{Nollert}H.-P.\ Nollert, 
{\sl Phys.\ Rev.\ }{\bf D47} (1993), 5253.
\bibitem{Motl} L.\ Motl,
{http://arXiv.org/abs/gr-qc/0212096}.
\bibitem{Dreyer} O.\ Dreyer, Phys. Rev. Lett. {\bf 90} (2003) 081301. 
\bibitem{MotlandNeitzke}L.\ Motl and A.\ Neitzke, 
{http://arXiv.org/abs/gr-qc/0301173}.
\bibitem{Corichi}A.\ Corichi, Phys. Rev. {\bf D67} (2003) 087502. 
\bibitem{Dirac} P. A. M. Dirac, ``The Principles of Quantum Mechanics (4th ed.)'',
Oxford University Press, 1958.
\bibitem{Jordan} P. Jordan, Zeitschrift f\"{u}r Physik, {\bf 44} (1927) 473.
\bibitem{Paulibook} I. Duck and E. C. G. Sudarshan, ``Pauli and the 
Spin-Statistics Theorem'', World Scientific, 1997.
\bibitem{Oeckl} R. Oeckl, J. Geom. Phys. {\bf 39} (2001) 233. 
\bibitem{LaidlawDeWitt} M. G. G. Laidlaw and C. Morette DeWitt, Phys. 
Rev. {\bf D3} (1971) 1375.
\bibitem{Donth} E. Donth, Physics Letters {\bf 32A} (1970) 209.
\bibitem{kinks} D. R. Finkelsetein and J. Rubensetin, J. Math. Phys. {\bf 9}
(1968) 1762.
\bibitem{Tscheuschner} R. Tscheuschner, Int. J. Theor. Phys. {\bf 28} (1989)
1269, erratum  Int. J. Theor. Phys. {\bf 29} (1990) 1437.
\bibitem{Balachandran} A. P. Balachandran, A. Daughton, Z. C. Gu, G. Marmo, R.
D. Sorkin and A. M. Srivastava, Mod. Phys. Lett. {\bf A5} (1990) 1575;
Int. J. Mod. Phys. {\bf A8} (1993) 2993.
\bibitem{Berry} M. V. Berry and J. M. Robbins, Proc. Roy. Soc. Lond. {\bf A453}
(1997) 1771; J. Phys. {\bf A33} (2000) L207. 
\bibitem{Geometricq} D. J. Simms and N. M. J. Woodhouse, ``Lectures
on Geometric Quantization'', Springer Lecture Notes in Physics 53,
Springer-Verlag, 1977; N. M. J. Woodhouse, ``Geometric Quantization (2nd ed.)'',
Oxford University Press, 1991; J.-M. Souriau, ``Structure of Dynamical
Systems'', Birkh\"{a}user Press, Boston, 1997.
\bibitem{Anastopoulos} C. Anastopoulos,  {http://arXiv.org/abs/quant-ph/0110169} .
\end{thebibliography}
\end{document}